\documentclass[reprint,twocolumn,superscriptaddress,pre,showpacs,aps]{revtex4-1}
\usepackage{amsmath,amssymb}
\usepackage{graphicx}
 \usepackage{color}
\usepackage{siunitx}
\usepackage{bm}

\newcommand{\ii}{\textrm{i}}
\newcommand{\ee}{\textrm{e}}

\newcommand{\dd}{\textrm{d}}

\newcommand{\inc}{\textrm{in}}
\newcommand{\sca}{\textrm{sc}}

\renewcommand{\vec}[1]{\bm{#1}}

\expandafter\ifx\csname package@font\endcsname\relax\else
 \expandafter\expandafter
 \expandafter\usepackage
 \expandafter\expandafter
 \expandafter{\csname package@font\endcsname}%
\fi

\begin{document}
\title[Radiation torque on a spheroid]
{Acoustic radiation torque exerted on a  subwavelength spheroidal particle by a traveling and standing plane wave}
\author{Jos\'e P. Le\~ao-Neto }
\affiliation{Campus Arapiraca/Unidade de Ensino Penedo, Universidade Federal de Alagoas, Penedo, Alagoas
	57200-000, Brazil}
\author{Jos\'e H. Lopes}
\affiliation{Grupo de F\'isica da Mat\'eria Condensada, N\'ucleo de Ci\^encias Exatas,
Universidade Federal de Alagoas,
Arapiraca, AL 57309-005, Brazil}
\author{Glauber T. Silva}
\email{gtomaz@fis.ufal.br}
\affiliation{Physical Acoustics Group,
Instituto de F\'isica,
Universidade Federal de Alagoas, 
Macei\'o, AL 57072-970, Brazil}

\date{\today}

\begin{abstract}
The nonlinear interaction of ultrasonic waves with a nonspherical particle 
may give rise to the acoustic radiation torque on the particle.
This phenomenon is investigated here considering a rigid prolate spheroidal particle of subwavelength dimensions
that is much smaller than the wavelength.
Using the partial wave expansion in spheroidal coordinates,
the radiation torque of a traveling and standing plane wave with arbitrary orientation
is exactly derived in the dipole approximation.
We obtain asymptotic expressions of the torque as the particle geometry approaches a sphere and a straight line.
As the particle is trapped in a pressure node of a standing plane wave, its radiation torque equals that of a traveling plane wave.
We also find how the torque changes with the particle aspect ratio.
Our findings are in excellent agreement with previous numerical computations.
Also, by analyzing the torque potential energy, we determine the stable and unstable spatial configuration available for a particle.
\end{abstract}

\maketitle

\section{Introduction}
There has been an increasing interest in 
studying the ultrasonic patterning of nonspherical particles 
such as fibers~\cite{Brodeur1990,Yamahira2000},
microrods~\cite{Saito1998},
 nanorods~\cite{Wang2012}, microfibers~\cite{Schwarz2015}, and stretched droplets~\cite{Foresti2013}.
These particles are translated by the action of 
the acoustic radiation force~\cite{Silva2014} and
may change orientation due to the acoustic radiation torque. Physically, this torque corresponds to 
the moment of the radiation stress on a particle~\cite{Zhang2011c}.

A typical example of the acoustic radiation torque phenomenon
on a nonspherical particle is the Rayleigh disk placed obliquely to the wave propagation direction~\cite{Rayleigh1945}.
Kotani~\cite{Kotani1933} analyzed the radiation torque on a Rayleigh disk caused by a traveling wave 
using oblate spheroidal wave functions.
King~\cite{King1935} used a cylindrical wave function basis to calculate the radiation  torque 
due to a standing plane wave.
Keller~\cite{Keller1957} obtained the radiation torque on infinitely long strips and disks employing the Babinet's principle.
Based on the angular momentum flux conservation, Maidanik~\cite{Maidanik1958} proposed a farfield calculation method and derived a solution for disks. 

Radiation torque solutions involving spheroidal particles are more scarce.
Some numerical methods have been employed to study this problem. 
The boundary element method (BEM) was utilized to calculate the torque exerted by a standing wave on a rigid spheroidal particle~\cite{Wijaya2015}.
The Born approximation with numerical quadrature was used
to obtain the radiation torque on a compressible spheroidal particle,
with density and compressibility close to those of the surrounding fluid.\cite{Jerome2019}
To date, the only analytical result to this problem was presented by Fan~\textit{et al.}~\cite{Fan2008}.
However, this investigation is mainly focused on developing a general theoretical framework
for arbitrarily shaped particles in the long-wavelength limit.

In this paper, we present the analytical solution of the radiation torque caused by a traveling and standing plane wave on a subwavelength spheroidal particle.
The incident waves may have arbitrary orientation regarding the particle major axis.
We obtain the radiation torque in the inviscid limit solving the corresponding scattering problem in spheroidal coordinates. 
The result is used in Maidanik's farfield method~\cite{Maidanik1958}. 
The inviscid approximation
is useful when 
the depth of the viscous boundary layer is much smaller than the particle size and streaming is weak.
We derive simple asymptotic expressions as the particle geometry approaches a sphere and straight line.
Excellent agreement is found between our method and BEM results considering a subwavelength prolate spheroid in a standing wave field~\cite{Wijaya2015}.
 
\section{Acoustic scattering}
\begin{figure}
	\centering
	\includegraphics[scale=.23]{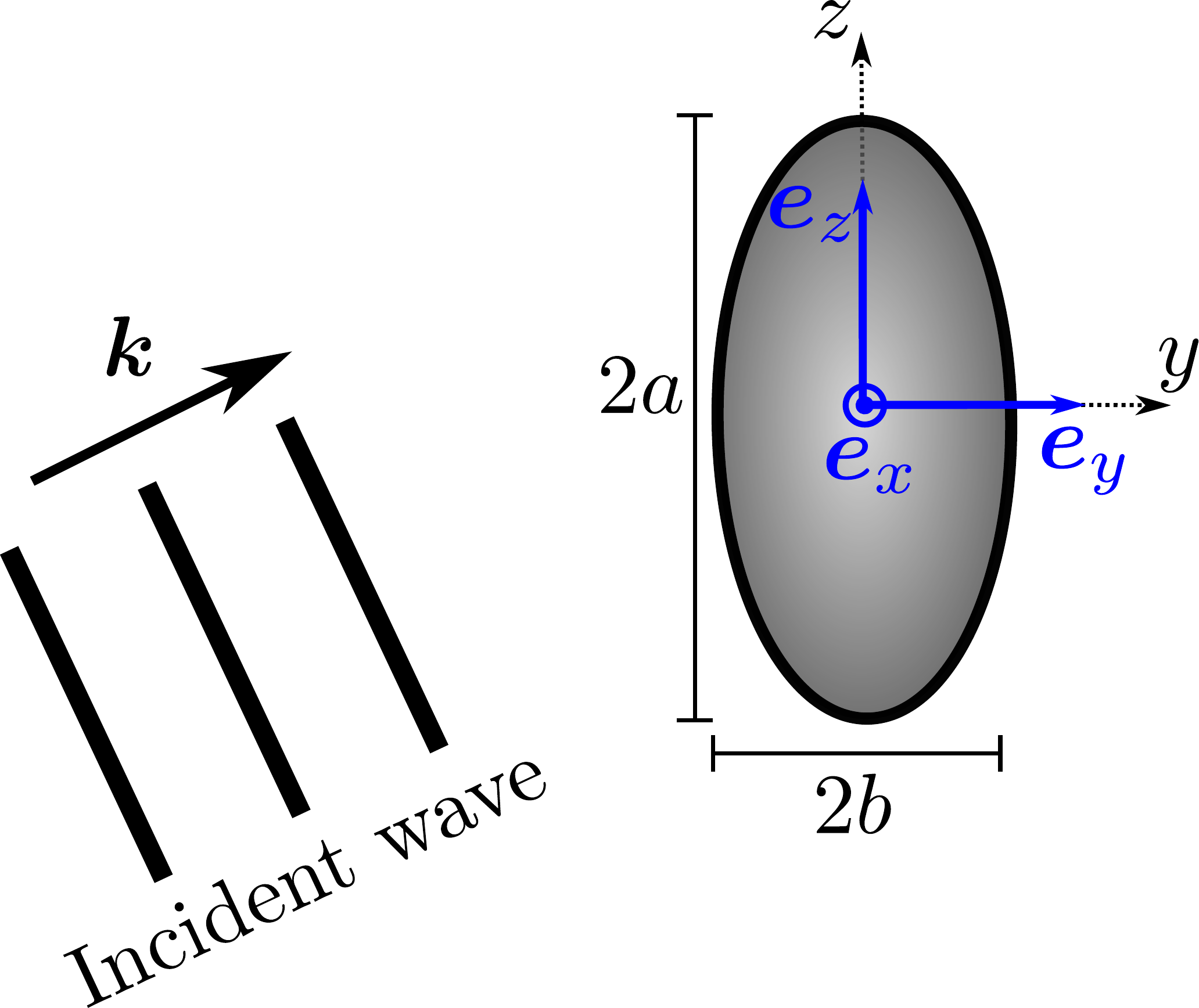}
	\caption{ 
		\label{fig:sketch}
		(color online)
		An incident wave with arbitrary wavevector $\vec{k}$ interacts with a prolate spheroid of major and minor axis denoted by $2a$ and $2b$, respectively.
		The origin of the Cartesian coordinate system is placed in the geometric center of the particle.
		The unit-vectors of the coordinate system are $\vec{e}_x$, $\vec{e}_y$, and $\vec{e}_z$.
	}
\end{figure}

Consider an inviscid fluid with density $\rho_0$, speed of sound $c_0$, and compressibility $\beta_0=1/\rho_0 c_0^2$.
A spheroidal particle  with a major and minor axis denoted by $2a$ and $2b$, respectively, 
is centered at the origin of the coordinate system--see Fig.~\ref{fig:sketch}.
The particle interfocal distance is $d=2 \sqrt{a^2-b^2}$.

A traveling or standing plane wave of angular frequency $\omega$ and wavenumber $k=2\pi/\lambda$, with $\lambda$ being the wavelength, is scattered by the particle.
For symmetry reasons, the acoustic scattering will be described in prolate spheroidal coordinates.
The Cartesian-to-spheroidal coordinate relations are
\begin{subequations}
	\begin{align}
	x &=\frac{d}{2}\sqrt{(\xi^2-1)(1-\eta^2)}\cos \varphi,\\
	y &=\frac{d}{2}\sqrt{(\xi^2-1)(1-\eta^2)}\sin \varphi,\\
	z &=\frac{d\xi \eta}{2},
	\end{align}
\end{subequations}
where 
$\xi\geq 1$ is radial distance,
$-1\leq\eta\leq1$, and  $0\leq\varphi\leq2\pi$ is azimuth angle. 
A prolate spheroidal particle corresponds to 
$\xi = \xi_0 = 2a/d$.
The aspect ratio of the particle is 
${a}/{b} = (1-\xi_0^{-2})^{-1/2}$,
while its volume is expressed by
$
V=4\pi ab^2/3=
\pi d^3 \xi_0 (\xi_0^2 -1)/6.
$

In the subwavelength scattering analysis, it is useful to define an expansion parameter 
in terms of the interfocal-to-wavelength ratio as
\begin{equation}
\label{kd}
\epsilon=\frac{kd}{2}\ll 1.
\end{equation}
In this limit, only the monopole and dipole modes of the incident and scattered waves are
needed to describe the particle-wave interaction.
Accordingly, the partial wave expansions of the incident and scattering potential velocities are given in prolate spheroidal coordinates by~\cite{Flammer2005}
\begin{subequations}
	\label{phi}
	\begin{align}
	\phi_\text{in} &= \phi_0 \sum_{n=0}^1 \sum_{m=-n}^n  a_{nm}S_{nm}(\epsilon,\eta) R_{nm}^{(1)}(\epsilon,\xi)\ee^{\ii m\varphi},
	\label{pin}\\
	\phi_\text{sc} &= \phi_0 \sum_{n=0}^1 \sum_{m=-n}^n  a_{nm}s_{nm}S_{nm}(\epsilon,\eta) R_{nm}^{(3)}(\epsilon,\xi)\ee^{\ii m\varphi},
	\label{psc}
	\end{align}
\end{subequations}
where  $S_{nm}$ is the angular function of the first kind, and $R_{nm}^{(1)}$ and $R_{nm}^{(3)}$  are the radial functions of the first and third kind, respectively.
The quantities $a_{nm}$ and $s_{nm}$ are the beam-shape and  scaled scattering coefficients.

Assuming that the particle behaves as a rigid and immovable spheroid, the normal component of fluid velocity on the particle surface satisfies $v_\xi(\xi_0)=\partial_{\xi}(\phi_\inc+\phi_\sca)_{\xi=\xi_0}=0$.
Using this condition into \eqref{phi} yields
the scattering coefficient as
\begin{equation}
s_{nm}=- \frac{\partial_{\xi} R_{nm}^{(1)}}{\partial_{\xi}R_{nm}^{(3)}}
\biggr|_{\xi=\xi_0}.
\label{coef:scattering}
\end{equation}

We shall see later that the acoustic radiation torque 
depends on the dipole moment ($n=1$) of the incident and scattered waves.
After Taylor-expanding 
the radial functions as given in \eqref{Rnms} around $\epsilon=0$ and use 
the result into \eqref{coef:scattering}, we obtain 
the dipole scattering coefficients as~\cite{Silva2018}
\begin{subequations}
	\label{scatt_spheroid}
	\begin{align}
	s_{10} &=\frac{\ii\epsilon^3}{6}f_{10}-\frac{\epsilon^6}{36}f_{10}^2 ,\\
	s_{1,-1}
	&=s_{11}=\frac{\ii\epsilon^3}{12}f_{11}-\frac{\epsilon^6}{144}f_{11}^2,
	\end{align}
\end{subequations}
where 
\begin{subequations}
	\label{factors}
	\begin{align}
	f_{10} &=\frac{2}{3}\left[\frac{\xi_0}{\xi_0^2-1}-\ln\left(\frac{\xi_0+1}{\sqrt{\xi_0^2-1}}\right)\right]^{-1} ,\\
	f_{11}
	&=\frac{8}{3}\left[\frac{2-\xi_0^2}{\xi_0(\xi_0^2-1)}+\ln\left(\frac{\xi_0+1}{\sqrt{\xi_0^2-1}}\right)\right]^{-1}
	\end{align}
\end{subequations}
are the scattering factors.

In the farfield $k\xi\gg 1$, 
the spheroidal wave functions in \eqref{phi} 
become spherical wave functions expressed in spherical coordinates $(r,\theta, \varphi)$ as follows~\cite{Silva2018}
\begin{subequations}
	\label{far}
	\begin{align}
	\phi_\inc &= \frac{\phi_0}{kr}
	\sum_{n=0}^1\sum_{m=-n}^n {a}_{nm}
	\sin\left(kr -\frac{n\pi }{2}\right)Y_n^m(\theta,\varphi),
	\\
	\phi_\sca &=\phi_0
	\frac{\ee^{\ii kr}}{kr} 
	\sum_{n=0}^1\sum_{m=-n}^n\ii^{-n-1}{a}_{nm}s_{nm}
	Y_n^m(\theta,\varphi),
	\end{align}
\end{subequations}
where 
$
Y_n^m(\theta,\varphi)$
is the spherical harmonic of $n$th-order and $m$th-degree.
Here the coefficient $a_{nm}$ describes an incident wave in spherical coordinates.
Some analytic expressions of beam-shape coefficients include  Bessel vortex and  Gaussian beams~\cite{Mitri2014}.
Numerical schemes and the addition theorem of spherical functions can be used to compute these coefficients 
for different types of beams~\cite{Silva2011a,Mitri2011,Silva2015a,Silva2013,Lopes2016}.

\section{Acoustic radiation torque}
\label{sec:ART}
The density of linear momentum flux conveyed by an acoustic wave is given by~\cite{Silva2011}
$\overline{\bf{P}} = -\overline{\mathcal{L}}{\bf I} + \rho_0 \overline{\vec{v}\vec{v}}$,
with the over bar denoting time-average over a wave period, and
$\bf I$ being the unit tensor.
The fields $\mathcal{L}$ and $ \rho_0 \vec{v}\vec{v}$ are the 
Lagrangian density and Reynolds' stress (a linear momentum flux).
The  density of angular momentum flux is defined as
$\overline{\bf{L}}=\vec{r}\times \overline{\bf{P}}=\vec{r}\times \rho_0\overline{\vec{v}\vec{v}}$, since $\vec{r}\times \bf{I} = \vec{0}$.
The acoustic radiation torque on a particle with surface $S_0$ is expressed by
\begin{equation}
\label{Ni}
\vec{\tau}_{\text{rad}} = 
\int_{S_0} 
\overline{ \bf{L}}\cdot \vec{n} \, \dd S
=
\int_{S_0} \left(\vec{r}\times
\rho_0 
\overline{ \vec{v} \vec{v}}\right)\cdot \vec{n} \, \dd S.
\end{equation}
The angular momentum flux satisfies the conservation law~\cite{Zhang2011c} $\nabla \cdot \overline{\bf{L}} = \vec{0}$.
Thus,
the integral in Eq.~\eqref{Ni} can be evaluated over a farfield virtual surface $S$ 
that encloses the particle.
Accordingly,  the radiation force is expressed by~\cite{Maidanik1958}
$
\vec{\tau}_{\text{rad}} = -\int_{S} \left(\vec{r}\times
\rho_0 
\overline{ \vec{v} \vec{v}}\right)\cdot \vec{n} \, \dd S. 
$

The fluid velocity is the sum of the velocity from the incident and scattered waves,
$\vec{v}=\vec{v}_\text{in} + \vec{v}_\text{sc}$.
Using this expression into 
the farfield radiation torque 
and noting that $\overline{\vec{v}\vec{v}}= (1/2)\text{Re}[\vec{v}\vec{v}^*]$,
we arrive at 
\begin{equation}
\label{Nend}
\vec{\tau}_{\text{rad}} = -\frac{\rho_0 r^2}{2}
\text{Re} \int_{\Omega_\text{s}}  
\vec{r}\times(\vec{v}_\text{in} \vec{v}_\text{sc}^*+ \vec{v}_\text{sc}\vec{v}_\text{in}^* + \vec{v}_\text{sc}\vec{v}_\text{sc}^*)\cdot \vec{e}_r\, \dd \Omega_\text{s},
\end{equation} 
where `Re' means the real part of, 
asterisk denotes complex conjugation, and $\Omega_\text{s}$ represents the unit-sphere.
In the inviscid approximation,
no torque is formed in the fluid without a particle; hence,
$\text{Re}\int_{\Omega_\text{s}}\vec{r} \times \vec{v}_\text{in} \vec{v}_\text{in}^*\cdot \vec{e}_r\,\dd\Omega_\text{s} = \vec{0}$.
Replacing the velocity potentials in \eqref{far} into Eq.~\eqref{Nend}, we find the radiation torque to the dipole approximation as~\cite{Silva2012}
\begin{subequations}
	\label{torque_rayleigh}
	\begin{align}
	\nonumber
	{\tau}_{\text{rad},x} &= - \frac{E_0 }{k^3\sqrt{2}}\,\text{Re} 
	\biggl[
	(a_{1,-1}+a_{11})(1+s_{11})a^*_{10}s^*_{10}\\
	&+
	a_{10}(1+s_{10})(a_{1,-1}^* + a_{11}^*)
	s_{11}^*
	\biggr],\\
	\nonumber
	{\tau}_{\text{rad},y} &=   \frac{E_0 }{k^3\sqrt{2}}\,\text{Re} \biggl[\ii\,
	(a_{1,-1}-a_{11})(1+s_{11})a^*_{10}s^*_{10}\\
	&-\ii\,
	a_{10}(1+s_{10})(a_{1,-1}^* -a_{11}^*)
	s_{11}^*
	\biggr],\\
	{\tau}_{\text{rad},z} &=  \frac{E_0 }{k^3 }\,\text{Re}
	\left[(|a_{1,-1}|^2 - |a_{11}|^2)
	(1+s_{11})s_{11}^*
	\right],
	\end{align}
\end{subequations}
where $E_0=\rho_0 k^2 \phi_0 /2$ is the characteristic energy density of the incident wave, with $p_0$ being its  peak pressure.

\section{Plane wave examples}
\subsection{Traveling plane wave}
The velocity potential of a traveling plane wave (TPW) propagating in an arbitrary direction is
\begin{equation}
\label{phiTPW}
\phi_\inc = \phi_0\, \ee^{\ii {\bm k}\cdot {\bm r} }.
\end{equation}
The wavevector reads
\begin{align}
\nonumber
{\bm k} &= k \vec{e}_k\\
&=
 k\left(\sin\theta_k \cos \varphi_k\, {\bm e}_x +
\sin\theta_k \sin \varphi_k \, {\bm e}_y +
\cos \theta_k \, {\bm e}_z\right).
\label{kplane}
\end{align} 
The angles $\theta_k$ and $\varphi_k$ are polar and azimuthal angles of the wave propagation direction.
Note that the orientation of the spheroidal particle is fixed along the direction determined by the unit-vector $\vec{e}_z$.
The orientation angle regarding the wave propagation direction $\vec{e}_k$ is determined from
$ \cos \theta_k =  \vec{e}_k \cdot \vec{e}_z$.

The beam-shape coefficient is obtained from
the TPW partial wave expansion in spherical coordinates
\begin{equation}
\label{standingwave2}
\textrm{e}^{\ii \bm{k}\cdot\bm{r}}=4\pi \sum_{n,m}\ii^n Y_n^{m*}(\theta_k,\varphi_k)\,j_n(kr)Y_n^{m}(\theta,\varphi).
\end{equation}
Thus, 
\begin{equation}
\label{anmPW}
{a}_{nm} =
4 \pi \ii^n Y_n^{m*}(\theta_k,\varphi_k).
\end{equation}
Replacing this coefficient into \eqref{torque_rayleigh}
and noting that $s_{1,-1} = s_{11}$,
we obtain
\begin{equation}
\vec{\tau}_\text{rad}^\text{T} = 
\frac{12\pi E_0}{k^3} 
\text{Im}
\left[
s_{10}^* + s_{11} +2 s_{10}^* s_{11}
\right](\vec{e}_k\cdot \vec{e}_z) \left(\vec{e}_k \times \vec{e}_z
\right),
\label{tauT}
\end{equation}
where we have used
 $\sin\theta_k \,(\sin\varphi_k\,\vec{e}_x 
-\cos\varphi_k\,\vec{e}_y)= (\vec{e}_k \times \vec{e}_z)$.
Inserting
the scattering coefficient
given in \eqref{scatt_spheroid} 
into Eq.~\eqref{tauT}
and noting that
$\epsilon^3=3k^3 V/[4\pi \xi_0(\xi_0^2-1)]$, 
we find the radiation torque as
\begin{subequations}
\begin{align}
\vec{\tau}^\text{T}_\text{rad} &= V E_0 
Q_\text{rad}
\left(\vec{e}_k\cdot \vec{e}_z\right) \left(\vec{e}_k \times \vec{e}_z
\right),
\label{NTPW}\\
Q_\text{rad} &=\frac{3}{4} \frac{f_{11} - 2 f_{10}}{ \xi_0(\xi_0^2 - 1)},
\label{Qrad}
\end{align}
\end{subequations}
where
$Q_\text{rad}$ is the radiation torque efficiency.
It is useful to define the characteristic radiation torque as
$\tau_0=V E_0 Q_\text{rad}$.

In the dipole approximation, the radiation torque does not involve self-interaction of the scattered wave.
The torque is  caused by the interference terms of the momentum flux
$\rho_0 \vec{v}_\text{in}\vec{v}_\text{sc}^*$
and $\rho_0 \vec{v}_\text{sc}\vec{v}_\text{in}^*$.

Due to  the axial symmetry of the particle, no radiation torque is produced  with end-on incidence ($\theta_k=0,\pi$).
It also vanishes in broadside incidence ($\theta_k=\pi/2$).
The maximum radiation torque is reached as $\theta_k=\pi/4$.
We also note that the radiation torque does not depend on frequency.

We may expand the torque efficiency in Eq.~\eqref{Qrad} as
the particle geometry approaches a sphere ($\xi_0\rightarrow \infty$),
\begin{equation}
\label{Qrad_sphere}
Q_\text{rad} =  \frac{9}{20}\left(
\frac{1}{\xi_0^2} + \frac{23}{70\xi_0^4}\right).
\end{equation}
The radiation torque vanishes for a spherical particle, $\lim_{\xi_0\rightarrow \infty}{Q}_\text{rad} = {0}$. 
This result is in agreement with the fact that no radiation torque is produced on a rigid sphere~\cite{Silva2012}.

The expansion of $Q_\text{rad}$
around $\xi_0=1$ gives  the radiation torque on a straight line.
Accordingly, we have
\begin{equation}
\label{Qrad_slender}
Q_\text{rad} =  
1 + 3(\xi_0-1)\left[2 + \ln\left(\frac{\xi_0-1}{2}\right)\right].
\end{equation}
For $\xi_0=1$, the radiation torque efficiency becomes $Q_\text{rad}=1$.

%
%
%

\subsection{Standing plane wave}
Consider a standing plane wave (SPW) formed by the superposition of two counter-propagating traveling
plane waves. The incident wave function is expressed by
\begin{align}
\nonumber
\phi_{\textrm{in}}&=\phi_0\cos\left[\bm{k}\cdot(\vec{r}+\vec{r}_\textrm{0})\right]\\
&=\phi_0\left[\textrm{e}^{\ii \bm{k}\cdot(\vec{r}+\vec{r}_\textrm{0})} +\textrm{e}^{-\ii \bm{k}\cdot(\vec{r}+\vec{r}_\textrm{0})}\right],
\label{standingwave}
\end{align}
where $\bm{r}_0$ points from the particle center to the nearest pressure antinode, which
lies in the same direction as the wavevector, $ \bm{k}\cdot\bm{r}_0=kr_0$. 
%

To obtain the  beam-shape coefficient of the SPW,  we use the TPW expansion from Eq.~(\ref{standingwave2}) into Eq.~(\ref{standingwave}) with the spherical harmonic  relation $Y_n^{m*}= (-1)^m Y_n^{-m}$.
Hence, 
\begin{equation}
{a}_{nm}=4\pi\cos\left(kr_0+\frac{n\pi}{2}\right)Y_n^{m*}(\theta_k,\varphi_k).
\end{equation}
Substituting this coefficient into the radiation torque components in~\eqref{torque_rayleigh}  yields
\begin{align}
\nonumber
\vec{\tau}_\text{rad}^\text{S} &= 
\frac{12\pi E_0}{k^3} 
\text{Im}
\left[
s_{10}^* + s_{11} +2 s_{10}^* s_{11}
\right]\\
&\times
\sin^2 k r_0\,(\vec{e}_k\cdot \vec{e}_z) \left(\vec{e}_k \times \vec{e}_z
\right).
\end{align}
Referring to Eq.~\eqref{NTPW}, the relation between the radiation torque of a standing and traveling plane wave is
given by
\begin{equation}
\label{NSPW}
\vec{\tau}_\text{rad}^\text{S} =\sin^2 k r_0\,
\vec{\tau}_\text{rad}^\text{T}.
\end{equation}
We note that Eqs.~\eqref{Qrad}, \eqref{Qrad_sphere}, and 
\eqref{Qrad_slender} are valid for a standing plane wave.

Due to the acoustic radiation force~\cite{Silva2018}, the spheroidal particle  has a tendency to be trapped in a pressure node $k r_0 =\pi/2$.
At the trapping point, 
the radiation torque of a standing plane wave is tantamount that of a traveling plane wave,
$\vec{\tau}_\text{rad}^\text{S} =
\vec{\tau}_\text{rad}^\text{T}.
$
%

\begin{figure}
	\centering
	\includegraphics[scale=.17]{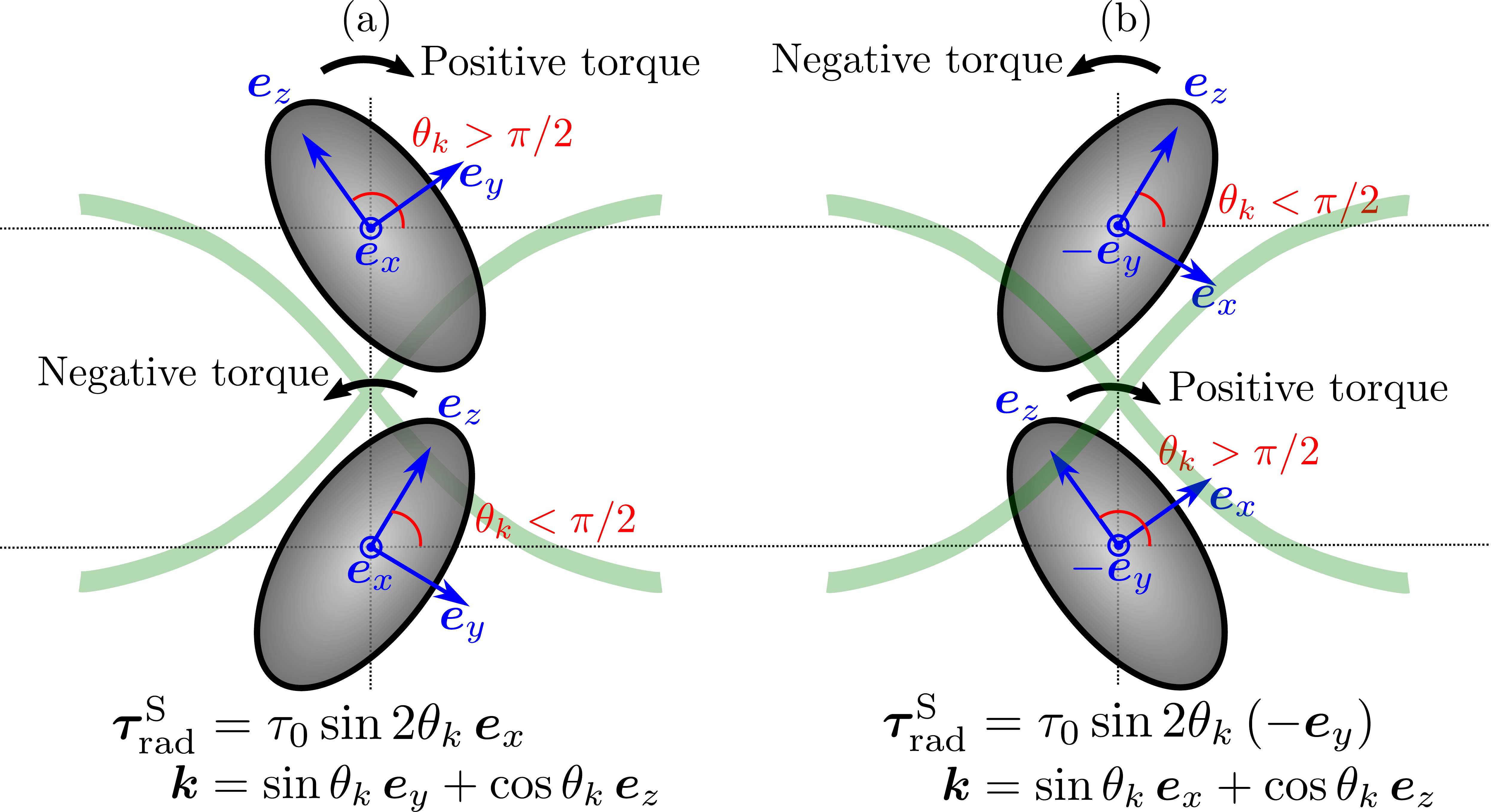}
	\caption{ 
		\label{fig:SPW}
		(color online)
		\small{ The radiation torque due to a standing plane wave (light green line) with wavevector in (a) the  $xz$ plane and (b) the $yz$ plane.
		In both cases, the radiation torque is positive  as $\theta_k<\pi/2$, and negative as $\theta_k>\pi/2$.
}}
\end{figure}

In Fig.~\ref{fig:SPW}, we depict a spheroidal particle
 under the influence of
the radiation torque of a SPW.
Panel (a) and (b) shows the SPW 
with the wavevectors 
$\vec{k}=k(\sin\theta_k\, \vec{e}_y+ \cos \theta_k\,\vec{e}_z)$ 
in the $yz$-plane
and
$\vec{k}=k(\sin\theta_k\, \vec{e}_x+ \cos \theta_k\,\vec{e}_z)$ in the $xz$-plane, respectively.
The corresponding radiation torques are
$
\vec{\tau}^\text{S}_\text{rad}=(\tau_0/2) \sin2\theta_k\, \vec{e}_x
$
and
$\vec{\tau}^\text{S}_\text{rad}=(\tau_0/2) \sin2\theta_k\, (-\vec{e}_y).$
Both torques set the particle to oscillate
around $\theta_k=\pi/2$, e.g., at right angle with the wave propagation direction.

\begin{figure}
	\centering
	\includegraphics[scale=.48]{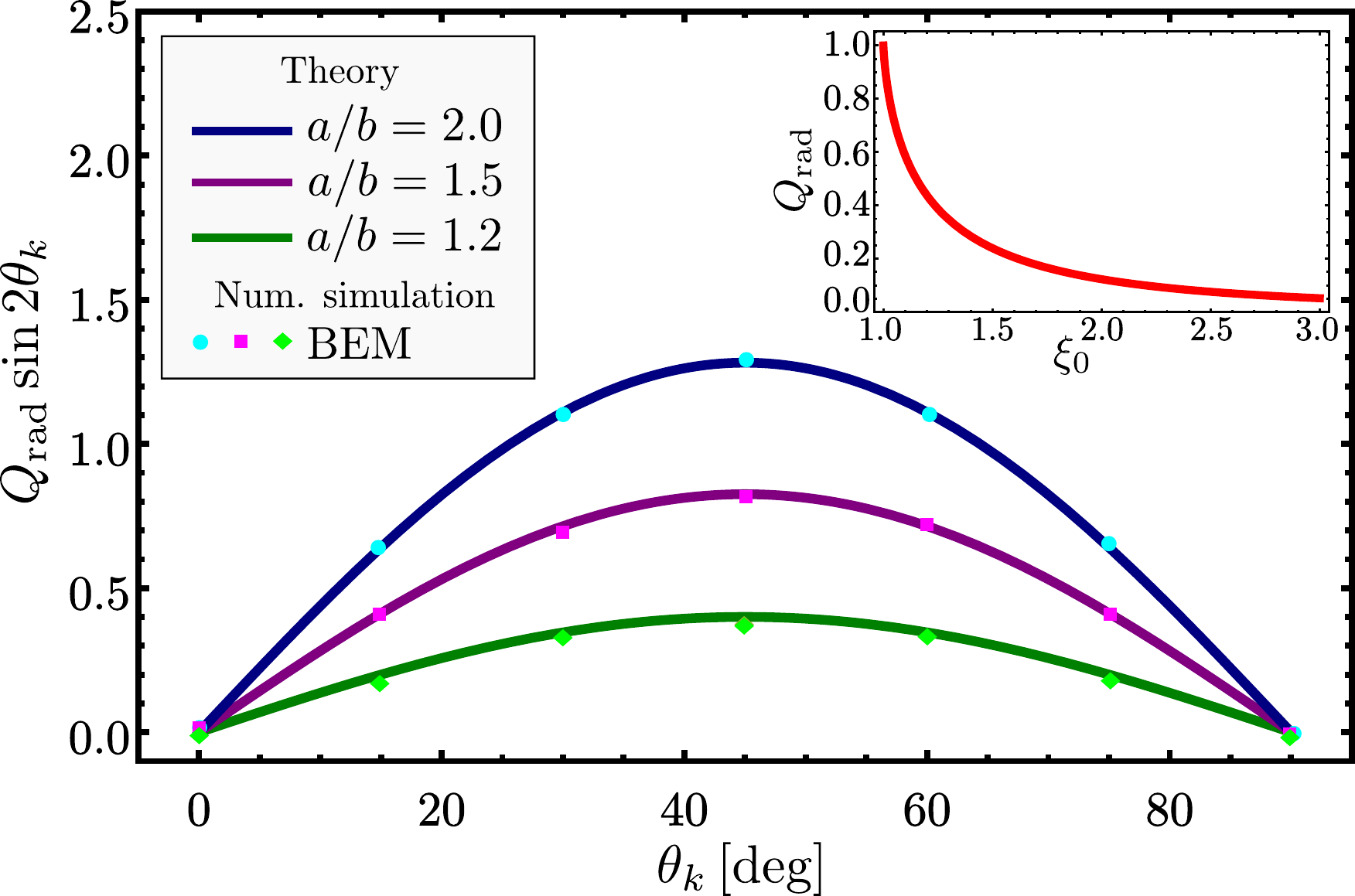}
	\caption{ 
		\label{fig:data}
		(color online)
	The radiation torque efficiency $Q_\text{rad}$ times $\sin 2\theta_k$ (solid lines) versus the incident angle $\theta_k$.
	The theoretical result is compared with numerical simulation data 
	considering particles with different aspect ratios.
	The data is obtained with the boundary element method (BEM)~\cite{Wijaya2015}. The inset shows $Q_\text{rad}$ as a function of the radial parameter $\xi_0$.
	}
\end{figure}

In Fig.~\ref{fig:data}, we compared the theoretical results with numerical simulation data from the boundary element method (BEM)~\cite{Wijaya2015}.
We consider the radiation torque efficiency $Q_\text{rad}$ times $\sin 2\theta_k$ of a standing plane wave and
particles with different aspect ratios, $a/b= 1.2$, $1.5$, $2.0$. 
The efficiency is compared with the dimensionless torque $T_\text{st}$ shown in \cite[Fig.~6b]{Wijaya2015}.
By direct inspection, we find that $Q_\text{rad}\sin 2\theta_k=4 T_\text{st}/k^3 V$, where 
$k^3 V = 0.00103745$.
Excellent agreement is found between our result and
numerical data.
In the inset, we note that the efficiency $Q_\text{rad}$
monotonically decreases with the radial parameter $\xi_0$
as it approaches to a spherical shape.

\section{Torque potential energy}

Moving on now to consider the potential energy of 
the radiation torque of a particle at a pressure node of
a standing plane wave.
This also corresponds to the case of a  traveling plane wave.
Assume that the radiation torque corresponds to the situation described in 
Fig.~\ref{fig:SPW}, panel (a).
The work done by the radiation torque from $\pi/2$ to an angle $\theta_k$ is
\begin{equation}
W = \int_{\pi/2}^{\theta_k} \tau^\text{S,T}_\text{rad}(\theta)\, \dd \theta=-\frac{\tau_0}{2}\cos^2\theta_k =-\frac{\tau_0}{2} \left(\vec{e}_k\cdot \vec{e}_z\right)^2.
\end{equation}
Therefore, 
the potential energy variation associated the radiation torque
is given by
\begin{equation}
\Delta U = U(\theta_k) - U(\pi/2) =  
-W =\frac{\tau_0}{2}\left(\vec{e}_k\cdot \vec{e}_z\right)^2.
\end{equation}
The minimum of the potential energy is $U(\pi/2)=0$, which shows 
the particle major axis has a tendency
to set itself broadside ($\theta_k=\pi/2$) on to the direction of the propagation of incident plane waves.
On the other hand, if the incidence angle
is $\theta_k=0$, the potential energy is maximum $U(0)=\tau_0/2$, which corresponds to an unstable equilibrium point.
This result agrees with 
experimental observations of polystyrene fibers~\cite{Yamahira2000}
and microrods~\cite{Saito1998}
that are trapped in a pressure node
of standing wave field.
These particles form a pattern in parallel alignment with the nodal planes.

\section{Discussion and conclusion}
We have presented analytic expressions of the acoustic radiation torque caused by a traveling and standing plane wave on a subwavelength spheroidal particle.
The results are exact to the dipole approximation of the
acoustic field expansions.
We found that the radiation torque is caused by interference between the incident and scattered waves.
Simple expressions of the asymptotic radiation torque as the particle geometry approaches a sphere and straight-line have also been derived.
The torque decreases monotonically as the particle assumes a spherical geometry.

When a particle is trapped in a pressure node of a standing wave, the radiation torque is the same as that of a traveling plane wave.
The peak radiation torque occurs as the wave incidence angle is $\theta_k=\pi/4$. 
The potential energy associated with the radiation torque reveals that the particle equilibrium orientation is broadside on ($\theta_k=\pi/2$) to the wave propagation direction.
Whereas, end-on incidence ($\theta_k=0$) promotes an unstable orientation setting the particle rotate toward the equilibrium orientation position ($\theta_k=\pi/2$).

The stable configuration predicted here 
agrees with previous experimental observation of fibers and microrods that are much smaller than the wavelength~\cite{Yamahira2000,Saito1998}.
Moreover, our findings are in excellent agreement with numerical simulation results based on the boundary element method~\cite{Wijaya2015}.
Finally, our results can be used to better analyze the dynamics of elongated cells and microorganisms (of prolate spheroidal shape) in acoustofluidic devices and of
reinforcing fibers in layered structures and in composite materials.

\begin{acknowledgments}
GTS thanks the National Council for Scientific and Technological
Development--CNPq, Brazil (Grant Nos. 401751/2016-3 and
307221/2016-4) for financial support.
\end{acknowledgments}

\vspace{1cm}

\appendix 
\section{Monopole and dipole radial functions}
\label{app:radfunctions}
In the long-wavelength approximation $\epsilon\ll1$,
the radial spheroidal functions can be expressed by~\cite{Burke1966}
\begin{widetext}
\begin{subequations}
	\label{Rnms}
\begin{align}
R_{00}^{(1)} &= 1 +
\frac{\epsilon^2}{18} \left(2-3 C_1^2\right) +\frac{\epsilon^4}{16200} \biggl[112-180 C_1^2 + 135 C_1^4
+\frac{\epsilon^2}{882} \left(2192 -8064 C_1^2 
+5670 C_1^4
-2835 C_1^6\right) \biggr],\\
R_{10}^{(1)} &=
\frac{\epsilon}{C_1}
+\frac{\epsilon^2C_1}{150}
\biggl[
2- 5C_1^2  + \frac{\epsilon^2}{4900}
\left(368-700C_1^2 + 875 C_1^4\right)
\biggr],\\
R_{11}^{(1)} &=
\frac{\epsilon S_1}{3} +
\frac{\epsilon^3S_1}{150}
\biggl[4 - 5 C_1^2 + \frac{\epsilon^2}{4900}(712-1400C_1^2 + 875 C_1^4)\biggr],
\\
\label{app:Rs}
R_{00}^{(2)} &=
-\frac{2}{\epsilon}
\biggl\{
L - \frac{\epsilon^2}{6}[6C_1+L(3C_2-5)]
+\frac{3}{5}\left(\frac{\epsilon}{6}\right)^4
\left[11 C_1 + 9 C_3+ \frac{L}{60}(1109 -1380 C_2 +135C_4)
\right]
\biggr\},\\
\nonumber
R_{10}^{(2)}&=
\frac{3 C_1}{\epsilon^2}
\biggl\{
2C_1 - \frac{C_2}{C_1} - 2L
-\left(\frac{\epsilon}{10}\right)^2 
\left[18C_1 - \frac{4C_2}{C_1} + L(22 - 10C_2)\right]
+ \frac{1}{882}\left(\frac{\epsilon}{10}\right)^4\\
&\times
\biggl[272313C1 - 864\frac{C_2}{C_1} + 7875 C_3 -
L(116073 - 99540 C_2 + 7875 C_2)\biggr]
\biggr\},
\\
\nonumber
R_{11}^{(2)} &=
-\frac{3S_1}{2 \epsilon^2}\biggl\{
\frac{C_1}{S_1^2} - 2L - \left(\frac{\epsilon}{10}\right)^4
\biggl[8C_1 \left(5-\frac{1}{S_1^2}\right) - 8L(33+5C_2)\biggr] 
-\frac{1}{196}\left(\frac{\epsilon}{10}\right)^2
\biggl[85800C_1 - 1750 C_3 
\\
&+  \frac{712 C_1}{S_1^2}
-L (106324 - 76950 C_2 - 1750 C_4)
\biggr]
\biggr\},\\
R_{nm}^{(3)}&= R_{nm}^{(1)} + \ii R_{nm}^{(2)},
\end{align}
\end{subequations}
\end{widetext}
\pagebreak

where $R_{nm}^{(2)}$ is the radial function of the second-kind.
We note that $R_{nm}^{(i)} = R^{(i)}_{n,-m}$, with $i=1,2,3$.
We also have
\begin{align}
\nonumber
C_n &= \frac{1}{2}\left[
(\sqrt{\xi^2 -1} + \xi)^n
+(\sqrt{\xi^2 -1} + \xi)^{-n}
\right],\\
\nonumber
S_n &= \frac{1}{2}\left[
(\sqrt{\xi^2 -1} + \xi)^n
-(\sqrt{\xi^2 -1} + \xi)^{-n}
\right],\\
\nonumber
L &= \frac{1}{2} \ln
\left(\frac{1 + (\sqrt{\xi^2 -1} + \xi)^{-1}}
{1 - (\sqrt{\xi^2 -1} + \xi)^{-1}}
\right).
\end{align}

%
%
%


\begin{thebibliography}{10}
	\def\enquote#1,{``#1,''}
	\expandafter\ifx\csname url\endcsname\relax
	\def\url#1{\texttt{#1}}\fi
	\expandafter\ifx\csname urlprefix\endcsname\relax\def\urlprefix{URL }\fi
	\providecommand{\bibinfo}[2]{#2}
	\def\plainquote#1{``#1''}
	\providecommand{\noopsort}[1]{}
	\providecommand{\switchargs}[2]{#2#1}
	\providecommand{\dourl}[1]{\href{http://#1}{\nolinkurl{#1}}}
	\providecommand{\dodoi}[1]{doi: \href{http://dx.doi.org/#1}{\nolinkurl{#1}}}
	\def\eatspace #1{#1}
	
	\bibitem{Brodeur1990}
	\bibinfo{author}{P.~Brodeur}, \enquote{\bibinfo{title}{Motion of
			fluid-suspended wave field}}, \bibinfo{journal}{Ultrasonics} \textbf{29},
	\bibinfo{pages}{302--307} (\bibinfo{year}{1990}).
	
	\bibitem{Yamahira2000}
	\bibinfo{author}{S.~Yamahira}, \bibinfo{author}{S.-I. Hanaka},
	\bibinfo{author}{M.~Kuwabara}, and \bibinfo{author}{S.~Asai},
	\enquote{\bibinfo{title}{Orientation of fibers in liquid by ultrasonic
			standing waves}}, \bibinfo{journal}{Jpn. J. Appl. Phys.} \textbf{39},
	\bibinfo{pages}{3683} (\bibinfo{year}{2000}).
	
	\bibitem{Saito1998}
	\bibinfo{author}{M.~Saito}, \bibinfo{author}{T.~Daian},
	\bibinfo{author}{K.~Hayashi}, and \bibinfo{author}{S.-Y. Izumida},
	\enquote{\bibinfo{title}{Fabrication of a polymer composite with periodic
			structure by the use of ultrasonic waves}}, \bibinfo{journal}{J. Appl. Phys.}
	\textbf{83}, \bibinfo{pages}{3490--3494} (\bibinfo{year}{1998}).
	
	\bibitem{Wang2012}
	\bibinfo{author}{W.~Wang}, \bibinfo{author}{L.~A. Castro},
	\bibinfo{author}{M.~Hoyos}, and \bibinfo{author}{T.~E. Mallouk},
	\enquote{\bibinfo{title}{Autonomous motion of metallic microrods propelled by
			ultrasound}}, \bibinfo{journal}{ACS Nano} \textbf{67},
	\bibinfo{pages}{6122--6132} (\bibinfo{year}{2012}).
	
	\bibitem{Schwarz2015}
	\bibinfo{author}{T.~Schwarz}, \bibinfo{author}{P.~Hahn},
	\bibinfo{author}{G.~Petit-Pierre}, and \bibinfo{author}{J.~Dual},
	\enquote{\bibinfo{title}{Rotation of fibers and other non-spherical particles
			by the acoustic radiation torque}}, \bibinfo{journal}{Microfluid Nanofluid}
	\textbf{18}, \bibinfo{pages}{65} (\bibinfo{year}{2015}).
	
	\bibitem{Foresti2013}
	\bibinfo{author}{D.~Foresti} and \bibinfo{author}{D.~Poulikakos},
	\enquote{\bibinfo{title}{{Acoustophoretic contactless elevation, orbital
				transport and spinning of matter in air}}}, \bibinfo{journal}{Phys. Rev.
		Lett.} \textbf{112}, \bibinfo{pages}{024301} (\bibinfo{year}{2014}).
	
	\bibitem{Silva2014}
	\bibinfo{author}{G.~T. Silva}, \enquote{\bibinfo{title}{{Acoustic radiation
				force and torque on an absorbing compressible particle in an inviscid
				fluid}}}, \bibinfo{journal}{J. Acoust. Soc. Am.} \textbf{136},
	\bibinfo{pages}{2405--2413} (\bibinfo{year}{2014}).
	
	\bibitem{Zhang2011c}
	\bibinfo{author}{L.~Zhang} and \bibinfo{author}{P.~L. Marston},
	\enquote{\bibinfo{title}{Acoustic radiation torque and the conservation of
			angular momentum (L)}}, \bibinfo{journal}{J. Acoust. Soc. Am.}
	\textbf{129}(4), \bibinfo{pages}{1679--1680} (\bibinfo{year}{2011}).
	
	\bibitem{Rayleigh1945}
	\bibinfo{author}{J.~W.~S. Rayleigh}, \emph{\bibinfo{title}{The Theory Of
			Sound}}, Vol.~\bibinfo{volume}{2}  (\bibinfo{publisher}{Dover Publications},
	\bibinfo{year}{1945}).
	
	\bibitem{Kotani1933}
	\bibinfo{author}{M.~Kotani}, \enquote{\bibinfo{title}{An acoustical problem
			relating to the theory of Rayleigh disc}}, \bibinfo{journal}{Proc. Phys.
		Math. Soc. Japan} \textbf{15}, \bibinfo{pages}{30} (\bibinfo{year}{1933}).
	
	\bibitem{King1935}
	\bibinfo{author}{L.~V. King}, \enquote{\bibinfo{title}{{On the theory of the
				inertia and diffraction corrections for the Rayleigh disc}}},
	\bibinfo{journal}{Proc. Royal Soc. A} \textbf{153}, \bibinfo{pages}{17}
	(\bibinfo{year}{1935}).
	
	\bibitem{Keller1957}
	\bibinfo{author}{J.~B. Keller}, \enquote{\bibinfo{title}{Acoustic torques and
			forces on disks}}, \bibinfo{journal}{J. Acoust. Soc. Am.} \textbf{29},
	\bibinfo{pages}{1085} (\bibinfo{year}{1957}).
	
	\bibitem{Maidanik1958}
	\bibinfo{author}{G.~Maidanik}, \enquote{\bibinfo{title}{{Torques due to
				acoustical radiation pressure}}}, \bibinfo{journal}{J. Acoust. Soc. Am.}
	\textbf{30}, \bibinfo{pages}{620--623} (\bibinfo{year}{1958}).
	
	\bibitem{Wijaya2015}
	\bibinfo{author}{F.~B. Wijaya} and \bibinfo{author}{K.-M. Lim},
	\enquote{\bibinfo{title}{Numerical calculation of acoustic radiation force
			and torque acting on rigid non-spherical particles}}, \bibinfo{journal}{Acta
		Acust. united Ac.} \textbf{101}, \bibinfo{pages}{531} (\bibinfo{year}{2015}).
	
	\bibitem{Jerome2019}
	\bibinfo{author}{T.~S. Jerome}, \bibinfo{author}{Y.~A. Ilinskii},
	\bibinfo{author}{E.~A. Zabolotskaya}, and \bibinfo{author}{M.~F. Hamilton},
	\enquote{\bibinfo{title}{Born approximation of acoustic radiation force and
			torque on soft objects of arbitrary shape}}, \bibinfo{journal}{J. Acoust.
		Soc. Am.} \textbf{145}, \bibinfo{pages}{36} (\bibinfo{year}{2019}).
	
	\bibitem{Fan2008}
	\bibinfo{author}{Z.~Fan}, \bibinfo{author}{D.~Mei}, \bibinfo{author}{K.~Yang},
	and \bibinfo{author}{Z.~Chen}, \enquote{\bibinfo{title}{Acoustic radiation
			torque on an irregularly shaped scatterer in an arbitrary sound field}},
	\bibinfo{journal}{J. Acoust. Soc. Am.} \textbf{124}(5),
	\bibinfo{pages}{2727--2732} (\bibinfo{year}{2008}).
	
	\bibitem{Flammer2005}
	\bibinfo{author}{C.~Flammer}, \emph{\bibinfo{title}{Spheroidal Wave Functions}}
	(\bibinfo{publisher}{Dover Publications}, \bibinfo{year}{2005}).
	
	\bibitem{Silva2018}
	\bibinfo{author}{G.~T. Silva} and \bibinfo{author}{B.~W. Drinkwater},
	\enquote{\bibinfo{title}{Acoustic radiation force exerted on a small
			spheroidal rigid particle by a beam of arbitrary wavefront: Examples of
			traveling and standing plane waves}}, \bibinfo{journal}{J. Acoustic. Soc.
		Am.} \textbf{144}, \bibinfo{pages}{EL453} (\bibinfo{year}{2018}).
	
	\bibitem{Mitri2014}
	\bibinfo{author}{F.~G. Mitri} and \bibinfo{author}{G.~T. Silva},
	\enquote{\bibinfo{title}{{Generalization of the extended optical theorem for
				scalar arbitrary-shape acoustical beams in spherical coordinates.}}},
	\bibinfo{journal}{Phys. Rev. E} \textbf{90}, \bibinfo{pages}{053204}
	(\bibinfo{year}{2014}).
	
	\bibitem{Silva2011a}
	\bibinfo{author}{G.~T. Silva}, \enquote{\bibinfo{title}{{Off-axis scattering of
				an ultrasound Bessel beam by a sphere}}}, \bibinfo{journal}{IEEE Trans.
		Ultrason. Ferroelectr. Freq. Control} \textbf{58}, \bibinfo{pages}{298--304}
	(\bibinfo{year}{2011}).
	
	\bibitem{Mitri2011}
	\bibinfo{author}{F.~G. Mitri} and \bibinfo{author}{G.~T. Silva},
	\enquote{\bibinfo{title}{{Off-axial acoustic scattering of a high-order
				Bessel vortex beam by a rigid sphere}}}, \bibinfo{journal}{Wave Motion}
	\textbf{46}, \bibinfo{pages}{392--400} (\bibinfo{year}{2011}).
	
	\bibitem{Silva2015a}
	\bibinfo{author}{G.~T. Silva}, \bibinfo{author}{A.~L. Baggio},
	\bibinfo{author}{J.~H. Lopes}, and \bibinfo{author}{F.~G. Mitri},
	\enquote{\bibinfo{title}{Computing the acoustic radiation force exerted on a
			sphere using the translational addition theorem}}, \bibinfo{journal}{IEEE
		Trans. Ultrason. Ferroelectr. Freq. Control} \textbf{62},
	\bibinfo{pages}{576--583} (\bibinfo{year}{2015}).
	
	\bibitem{Silva2013}
	\bibinfo{author}{G.~T. Silva}, \bibinfo{author}{J.~H. Lopes}, and
	\bibinfo{author}{F.~G. Mitri}, \enquote{\bibinfo{title}{{Off-axial acoustic
				radiation force of repulsor and tractor bessel beams on a sphere.}}},
	\bibinfo{journal}{IEEE Trans. Ultrason. Ferroelectr. Freq. Control}
	\textbf{60}, \bibinfo{pages}{1207--1212} (\bibinfo{year}{2013}).
	
	\bibitem{Lopes2016}
	\bibinfo{author}{J.~H. Lopes}, \bibinfo{author}{M.~Azarpeyvand}, and
	\bibinfo{author}{G.~T. Silva}, \enquote{\bibinfo{title}{Acoustic interaction
			forces and torques acting on suspended spheres in an ideal fluid}},
	\bibinfo{journal}{IEEE Trans. Ultrason. Ferroelectr. Freq. Control}
	\textbf{63}, \bibinfo{pages}{186--97} (\bibinfo{year}{2016}).
	
	\bibitem{Silva2011}
	\bibinfo{author}{G.~T. Silva}, \enquote{\bibinfo{title}{{An expression for the
				radiation force exerted by an acoustic beam with arbitrary wavefront}}},
	\bibinfo{journal}{J. Acoust. Soc. Am.} \textbf{130},
	\bibinfo{pages}{3541--3545} (\bibinfo{year}{2011}).
	
	\bibitem{Silva2012}
	\bibinfo{author}{G.~T. Silva}, \bibinfo{author}{T.~P. Lobo}, and
	\bibinfo{author}{F.~G. Mitri}, \enquote{\bibinfo{title}{Radiation torque
			produced by an arbitrary acoustic wave}}, \bibinfo{journal}{Europhys. Lett.}
	\textbf{97}(5), \bibinfo{pages}{54003} (\bibinfo{year}{2012}).
	
	\bibitem{Burke1966}
	\bibinfo{author}{J.~E. Burke}, \enquote{\bibinfo{title}{Note on spheroidal wave
			functions}}, \bibinfo{journal}{Stud. Appl. Math.} \textbf{45},
	\bibinfo{pages}{425--431} (\bibinfo{year}{1966}).
	
\end{thebibliography}

\end{document}